\documentclass[11pt,preprint]{aastex}
\usepackage{graphicx}

\slugcomment{}

\shorttitle{The NIR Ca{\sc{ii}} triplet indices}
\shortauthors{W. Du et al.}

\begin{document}


\title{A new synthetic library of the Near-Infrared Ca{\sc{ii}} triplet indices. I.Index Definition, Calibration and Relations with stellar atmospheric parameters}


\author{
W.\ Du\altaffilmark{1,2,3,4},
A L.\ Luo\altaffilmark{1,2,5} and
Y. H.\ Zhao\altaffilmark{1,2}}
\affil{$^1$ National Astronomical Observatories, Chinese Academy of Sciences, Beijing, 100012, China}
\affil{$^2$ Key Laboratory of Optical Astronomy, NAOC, Beijing, 100012, China}
\affil{$^3$ Graduate University of Chinese Academy of Sciences, Beijing, 100049, China}
\affil{$^4$ email: {\texttt wdu@nao.cas.cn}}
\affil{$^5$ email: {\texttt lal@nao.cas.cn}}




\begin{abstract}
Adopting the SPECTRUM package which is a stellar spectral synthesis program, we have synthesized a comprehensive set of 2,890 Near-InfraRed (NIR) synthetic spectra with a resolution and wavelength sampling similar to the SDSS and the forthcoming LAMOST spectra. During the synthesis, we have applied the `New grids of ATLAS9 Model Atmosphere' to provide a grid of local thermodynamic equilibrium (LTE) model atmospheres for effective temperatures ($T_{eff}$) ranging from 3,500 to 7,500 K, for surface gravities (log $g$) from 0.5 to 5.0 dex, for metallicities ([Fe/H]) from -4.0 to 0.5 dex, for abundances solar ([$\alpha$/Fe] = 0.0 dex) and non-solar ([$\alpha$/Fe] = +0.4 dex). So this synthetic stellar library is composed of 1,350 solor scaled abundance (SSA) and 1,530 non-solar scaled abundance (NSSA) spectra, grounding on which we have defined a new set of NIR Ca{\sc{ii}} triplet indices and an index CaT as the sum of the three. Then, these defined indices have been automatically measured on every spectrum of the synthetic stellar library and calibrated with the indices computed on the observational spectra from the INDO-U.S. stellar library. In order to check the effect of $\alpha$-element  enhancement on the so-defined Ca{\sc{ii}} indices, we have compared indices measured on the SSA spectra with those on the NSSA ones at the same terns of stellar parameters ($T_{eff}$, log $g$, [Fe/H]), and luckily, little influences of $\alpha$-element enhancement has been found. Furthermore, comparisons of our synthetic indices with the observational ones from measurements on the INDO-U.S. stellar library, the SDSS-DR7, and SDSS-DR8 spectroscopic survey have been presented respectively for dwarfs and giants in specific. For dwarfs, our synthetic indices could well reproduce the behaviours of the observational indices versus stellar parameters, which verifies the validity of our index definitions for dwarfs. While for giants, the consistency between our synthetic indices and the observational ones does not appear that good. However, a new synthetic library of NIR Ca{\sc{ii}} indices has been founded for deeper studies on the NIR waveband of stellar spectra, and this library is particularly appropriate for the SDSS and the forthcoming LAMOST stellar spectra. Later on, we have regressed the strength of the CaT index as a function of stellar parameters for both dwarfs and giants after a series of experimental investigations into relations of the indices with stellar parameters. For dwarfs, log $g$ has little effect on the indices while [Fe/H] and $T_{eff}$ play a role together and the leading factor is probably [Fe/H] which changes the stength of the indices by a positive trend. For giants, log $g$ starts to influence on the strength of the indices by a negative trend for the metal-poor and even impact deeply for the metal-rich, and besides, [Fe/H] and $T_{eff}$ still matter. Ultimately, a supplemental experimentation has been carried out to show that spectral noises do have effects on our set of NIR Ca{\sc{ii}} indices. However, the infection is weak enough to be ignored if the signal-to-noise ratio (SNR) falls below 20. 

\end{abstract}





\section{Introduction}

Due to more and more spectroscopic surveys, a huger number of spectra are turning available for studies on the physical properties (atmospheric parameters of stars; abundances, composition and ages of composite stellar populations of galaxies) by extracting information hidden in the absorption lines. However, problems of the limited sampling of spectral features caused by the detector pixel and the signal-to-noise ratio (SNR) seriously hamper the detailed fitting of the weak spectral features. So an alternative option is expected to extract physical information from the spectrum with no need of fitting faint absorption lines. Therefore, the concept of ``spectral indices" has been introduced. This good idea for spectral analysis is to define a set of ``spectral indices" each of which is an estimate of the equivalent width of a spectral feature measured by taking the ratio of the flux of feature in the central band to that of pseudo-continuum determined by two adjacent continuum bands. So far, various definitions for absorption features have existed as different spectral index systems of which the most successful is the Lick/IDS index system \citep{b5, b10, b11, b12, b13, b14} which defines a set of 25 spectral indices in the optical waveband. Although it is widely used, Lick/IDS index system lacks for definitions of absorption features in the Near-InfraRed (NIR) waveband since the Lick spectra only cover from 4000 $\AA$ to 6400 $\AA$ in wavelength. 

In the NIR region ($\lambda\lambda$ 7900 $\sim$ 9100 $\AA$) of stellar spectra of types A $\sim$ M, the strongest atomic feature is definitely Ca{\sc{ii}} triplet ($\lambda\lambda$ 8498, 8542, 8662 $\AA$) which offers several advantages from the observational point of view for being the strongest metal line in the NIR spectrum, exceeding Na{\sc{i}} ($\lambda$8190) in strength by at least a factor of 5, and lying in a region relatively free of atmospheric absorptions. The Ca{\sc{ii}} triplet has been used as a luminosity indicator \citep{b4}, and its strength is simultaneously a good metallicity indicator in stellar systems where the NIR light is dominated by stars of a restricted range of surface gravities \citep{b1}. Beside, it has provided a potential tool to study the properties of stellar populations since the old stellar populations emit most of their light in this wavelength region \citep{b1}. 

Progresses have been previously made at studies on the NIR Ca{\sc{ii}} triplet. By studying a sample of 62 stars of types F0 $\sim$ M5, \citet{b4} has derived that the equivalent width (EW) of the NIR Ca{\sc{ii}} triplet only correlates strongly with log $g$ in stars. This single-valued correlation has later been confirmed by \citet{b15}. However, by a set of definitions for the Ca{\sc{ii}} indices on a sample of 106 late-type stars, \citet{b1} has demonstrated that in the high-metallicity range, the strength of the indices merely depends on log $g$ while in the low-metallicity range, [Fe/H] turns to play a leading role. Afterwards, by refined definitions for the Ca{\sc{ii}} indices on a larger number of stars with a higher resolution, \citet{b16} has clarified that $T_{eff}$ has an important impact factor for the low $T_{eff}$ stars and log $g$ plays a more important role for giants than dwarfs. \citet{b2} has quantitatively estimated the strength of the Ca{\sc{ii}} indices by introducing a set of ``generic indices".

Although several works have been made on definitions for the NIR Ca{\sc{ii}} triplet indices\citep{b4,b1,b2,b16}, they are all based on samples of less stars with higher resolution and denser wavelength sampling than the Sloan Digital Sky Survey Data Release7 (SDSS DR7; \citet{b18,b19}) and the forthcoming Large Sky Area Multi-Oject Fiber Spectroscopic Telescope (LAMOST; \citet{b20}) spectra in the 8300$\AA$ $\sim$ 8900$\AA$ region. Besides, those stars also have inaccurate estimations for stellar parameters. In order to cope with these disadvantages, we are expecting to establish a new index system of the Ca{\sc{ii}} triplet which would be the most appropriate and convenient for use on the SDSS and the forthcoming LAMOST NIR spectra. We will perform this work on on a larger number of synthetic stellar spectra with approximately the same spectral resolution and wavelength sampling as the SDSS-DR7 and the forthcoming LAMOST spectra. 

The aim of this work is to constitute a new synthetic library of a set of the NIR Ca{\sc{ii}} indices for deeper studies on the NIR waveband of spectra, especially of the SDSS and the forthcoming LAMOST stellar spectra. This library could also be a supplemental part of the NIR indices to the work of \citet{b3} which has established a synthetic library of several optical indices for SDSS and the forthcoming LAMOST spectra. In Section 2, we generate a large number of synthetic stellar spectra with various stellar parameters, on which a new set of the NIR Ca{\sc{ii}} indices has been defined and measured. Then, we also investigate the dependence of the Ca{\sc{ii}} indices on $\alpha$-element enhancement. In Section 3, we calibrate our synthetic indices with the observational indices measured on the INDO-U.S. stars \citep{b6}. Then, we compare the calibrated synthetic Ca{\sc{ii}} indices with the observational INDO-U.S indices. Besides, tests of the calibrated synthetic indices on SDSS-DR7 and SDSS-DR8 \citep{b21} stars have been carried out. In Section 4, we analyse the relations of the calibrated synthetic Ca{\sc{ii}} indices with stellar parameters and approximately estimate the stength of the CaT index by stellar parameters. We compare the Ca{\sc{ii}} triplet indices with the CaHK index in Section 5. Finally, we summarize this paper in Section 6. As a supplement, we have performed a series of experiments in Appendix A to check into the effect of SNR on Ca{\sc{ii}} indices, and found that such an effect could be ignored until the SNR falls below 20.

\section{The Synthetic Indices}


\subsection[]{Generate synthetic stellar spectra}
Above all, a saying is crucial to be stressed that it is not simple to model Ca{\sc{ii}} triplet lines using the local thermodynamic equilibrium (LTE) models since the line cores are formed in the stellar chromosphere and, hence, non-local thermodynamic equilibrium (NLTE) models are required. So \citet{b17} has carried out, using NLTE models for the first time, the computation of the line cores, deriving full equivalent-widths and they have shown, however, that the effects of departures from LTE are negligible since the equivalent widths are dominated by the line wings. Considering that our Ca{\sc{ii}} triplet indices are estimated as equivalent widths, we could use LTE models for spectral synthesis without bringing departures in. Therefore, we compute the LTE model synthetic stellar spectra as our synthetic sample in this work, using the stellar spectral synthesis program SPECTRUM \citep{b22}. While using the SPECTRUM package, we set the wavelength sampling to be 0.05 ($\Delta$$\lambda$ = 0.05), rotational velocity to be 0 km/s, microturbulent velocity to be 2 km/s (e = 2 km/s), and pick the `` New Grids of ATLAS9 Model Atmospheres" \citep{b24} as the stellar atmosphere models from which we extract a subgrid as follows.  

Effective temperature($T_{eff}$): 3,500 $\sim$ 7,500 K with a step of 250 K

Surface gravity(log $g$): 0.5 $\sim$ 5.0 dex with a step of 0.5 dex

For chemical composition, there are two assumptions which lead to different subgrids. One is the solar scaled abundances (SSA) grid ([$\alpha$/Fe] = 0.0) which has 8 points for metallicity ([Fe/H] = -2.5, -2.0, -1.5, -1.0, -0.5, 0.0, 0.2, and 0.5 dex) and the other is the non-solar scaled abundances (NSSA) grid with [$\alpha$/Fe] = 0.4 dex which has 9 points for metallicity ([Fe/H] = -4.0, -2.5, -2.0, -1.5, -1.0, -0.5, 0.0, 0.2, and 0.5 dex). After spectral synthesis of the above grids, we degrade the resolution of the synthetic spectra to R = 1800 (FWHM $\approxeq$ 4.8 $\AA$ in $\lambda\lambda$8300 $\sim$ 8900 $\AA$) and change the wavelength spacing to be 2.0 $\AA$ to match with SDSS and the forthcoming LAMOST spectra by using the SMOOTH2 task of the SPECTRUM package which convolves a Gaussian linespread function with the synthetic spectrum to smooth the spectrum to a desired resolution and then output the smoothed spectrum with a desired wavelength spacing. Hence, we finally derive 2890 synthetic spectra composed of 1360 SSA model and 1530 NSSA model ones in the wavelength range of 8300$\AA$~8900$\AA$. It is necessary to be emphasized that the shapes of the synthetic spectra of hotter stars ($T_{eff}$ $>$ 6000 K) appear a little abnormal for some reason so we have to choose $T_{eff}$ = 6000 K as a demarcation. We call the synthetic spectra with $T_{eff}$ $\leqslant$ 6000 K the Large Synthetic Sample (LSS) and those with 6000 K $\leqslant$ $T_{eff}$ $\leqslant$ 7500 K the Small Synthetic Sample (SSS). The LSS has 1870 spectra composed of 880 SSA model spectra and 990 NSSA model spectra while the SSS has 1020 spectra consist of 480 SSA and 540 NSSA ones.

\subsection[]{Definition and measurement of the Ca{\sc{ii}} triplet indices}
An absorption strength is always expressed in terms of ``an index ", which is usually composed of a ``feature" bandpass and the flanking blue and red ``continuum" bandpasses.
The bandpasses should be defined by the following criteria: the ``feature" bandpass needs to be defined to center on the feature of interest, the ``continuum" bandpasses require to be located near the feature bandpass, to be in regions of less blended absorptions, and need for relative insensitivity to stellar velocity dispersion broadening. This last mandates a minimum length for bandpass definitions \citep{b5}.

As is known, several atomic lines of intermediate strength (Fe{\sc{i}} $\lambda\lambda$8514.1, 8674.8, 8688.6, 8824.2$\AA$; Mg{\sc{i}} $\lambda$8806.8$\AA$ and Ti{\sc{i}} $\lambda$8435.0$\AA$) exist in the wavelength region of the Ca{\sc{ii}} triplet lines. Moreover, the hydrogen Paschen series ($\lambda\lambda$8359.0, 8374.4, 8392.4, 8413.3, 8438.0, 8467.3, 8502.5, 8545.4, 8598.4, 8665.0, 9750.5, 8862.8, from P19 to P10, respectively) are also blended into this wavelength region. However, the Ca{\sc{ii}} triplet absorptions are fortunately the strongest in stellar spectra of types F5 $\sim$ M2, and much stronger than other atomic lines \citep{b2}. Simultaneously, the Paschen series nearly disappear in the spectra of stars colder than 6,000 K and thus could affect little on the Ca{\sc{ii}} triplet (see details in Figure 4 in \citet{b2}).

To get a general idea, we check the centers, shapes, wings, and the local continuum of the Ca{\sc{ii}} triplet on the synthetic spectra of our LSS carefully. Then, we make definitions for the Ca{\sc{ii}} indices by choosing the central and continuum regions to cover the feature as complete as possible and simultaneously minimize the effect of the atomic or Paschen lines. By trial and error, we make a final definition for the Ca{\sc{ii}} triplet indices. The interval specifications for each line index are tabulated in Table 1. Like the Lick/IDS index system, each line index is measured by the ratio of the flux contained in a wavelength region centered on the feature relative to a pseudo local continuum represented by a straight line between the midpoints of the blue and red ``continuum" regions close to the feature (for details see \citet{b11}). Such a line-strength index whose definition is so similar to equivalent width gives a quantitative measure of a spectral signature. Additionally, we define a new compound index CaT as the sum of the three single index.
\begin{table*}
\centering
  \caption{Wavelength intervals for each line index of the NIR Ca{\sc{ii} triplet}
            }
  \begin{tabular*}{120mm}{@{}cccc}
\hline
  Identification & Blue               & Line              &Red  \\
                 & Bandpass($\AA$)  & Bandpass($\AA$) &Bandpass($\AA$)\\
 \hline
Ca{\sc{ii}}$\lambda$8498$\AA$	  	&8470.0$\sim$8490.0	&8491.0$\sim$8508.0      &8521.0$\sim$8531.0	\\
Ca{\sc{ii}}$\lambda$8542$\AA$	  	&8521.0$\sim$8531.0	&8532.0$\sim$8553.0	 &8563.0$\sim$8575.0\\
Ca{\sc{ii}}$\lambda$8662$\AA$	  	&8643.0$\sim$8653.0	&8655.0$\sim$8670.0	 &8695.0$\sim$8710.0\\

\hline
 
\end{tabular*}
\end{table*}

According to the definitions, we automatically compute the Ca{\sc{ii}} indices on the LSS and SSS. These defined indices would as well be appropriate for studies of the NIR waveband of SDSS and the forthcoming LAMOST spectra. Conveniently, no consideration of matching resolutions is needed before using these indices on the SDSS and the forthcoming LAMOST spectra.

\subsection[]{Effect of $\alpha$-element enhancement}
In this paper we investigate the effect of $\alpha$-element enhancement on the Ca{\sc{ii}} indices. As our SSA model spectra do not have any $\alpha$-element enhancement ([$\alpha/Fe$] = 0.0) while the NSSA ones do ([$\alpha/Fe$] = 0.4), we could check the effect directly via comparisons between the indices of the SSA spectra with those of the NSSA ones at the same terns of stellar parameters ($T_{eff}$, log $g$, [Fe/H]). For a clarification, we show the comparisons for each index in Figure 1 where it is obviously stated that the Ca{\sc{ii}} indices do not show any significant differences between SSA and NSSA indices. Furthermore, it seems that the consistency between the indices of SSA and NSSA is much better for the SSS (the red points) than the LSS (the black points); however, we can not be sure of the result for SSS because the spectra of SSS might not be correct from synthesis as is mentioned in Section 2.1. So we just believe that the NIR Ca{\sc{ii}} indices are all quasi-independent on $\alpha$-element enhancement for stars cooler than 6000 K. 
\begin{figure*} 
\centering
 \includegraphics[scale=1.6]{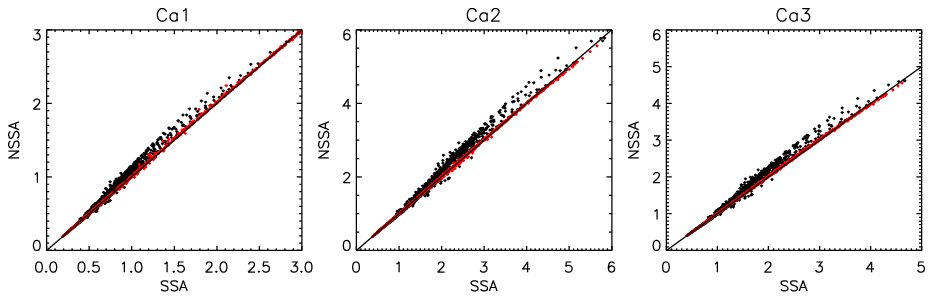}
 \caption[]{The effect of $\alpha$-element enhancement on the Ca{\sc{ii}} triplet indices. The x-axis presents index values of the SSA spectra ([$\alpha$/Fe] = 0.0) while the y-axis shows those of the NSSA spectra ([$\alpha$/Fe] = +0.4). The black and red points stand for the LSS and SSS spectra, respectively. }
\end{figure*}

\section[]{Calibration of synthetic indices}
In order to calibrate our synthetic Ca{\sc{ii}} indices, we are required to compare them with those measured on reference stars with known and accurate atmospheric parameters ($T_{eff}$, log $g$, [Fe/H]). Since the INDO-U.S. stellar library \citep{b6} covers the region of the NIR Ca{\sc{ii}} triplet in wavelength, we pick up stars with stellar parameters within our subgrid from the INDO-U.S. library as the test sample which includes 687 stars. Moreover, 564 (138 dwarfs and 426 giants) out of the 687 stars are cooler than 6000 K, so we select these 564 INDO-U.S. stars as the calibration sample. Besides, the stellar parameters for the test sample we choose are from the version of  \citet{b23}, which has estimated parameters more accurately for the complete INDO-U.S. stars, particularly for giants (see details in \citet{b23}). After degrading the resolution of the calibration sample to R = 1800, the 564 indices are computed on the resolution-matched sample. Here, we consider these 564 observational indices as the Observational Indices Sample (OIS). Then, we calculate the corresponding 564 synthetic indices with the same terns of stellar parameters as those of the calibration sample by linear interpolations in our SSA parameters-index grid ($T_{eff}$, [Fe/H], log $g$, index). Here, we regard these 564 synthetic indices as the Synthetic Indices Sample (SIS). By performing linear regressions between OIS and SIS, we derive a set of transformation coefficients (slopes and constants). Expectedly, the slopes of the linear fits are all close to one, which gives us confidence on the consistency between theoretical and observational spectra. However, the constant factors for dwarfs differs from those for giants. In detail, the constant calibration factors are respectively 0.196, 0.645 and 0.425 for Ca1, Ca2, and Ca3 index for dwarfs and 0.340, 0.719 and 0.555 for giants. As an example of calibration, we show the calibrations of the three indices for dwarfs in Figure 2 where the black lines represent the calibration lines which could transform our synthetic indices into observational ones. Similarly, the transformations for Ca1, Ca2, and Ca3 index are respectively Y = 1.238X + 0.340, Y = 1.258X + 0.719, and Y = 1.148X + 0.555 for giants. As the Ca{\sc{ii}} indices are all quasi-independent on $\alpha$-element enhancement (See details in Section 2.3), we do not repeat the calibration process for NSSA synthetic indices. Instead, the derived sets of transformation coefficients are used to convert both the SSA and NSSA synthetic indices into the observational indices system.
\begin{figure*}
\centering
 \includegraphics[scale=1.7]{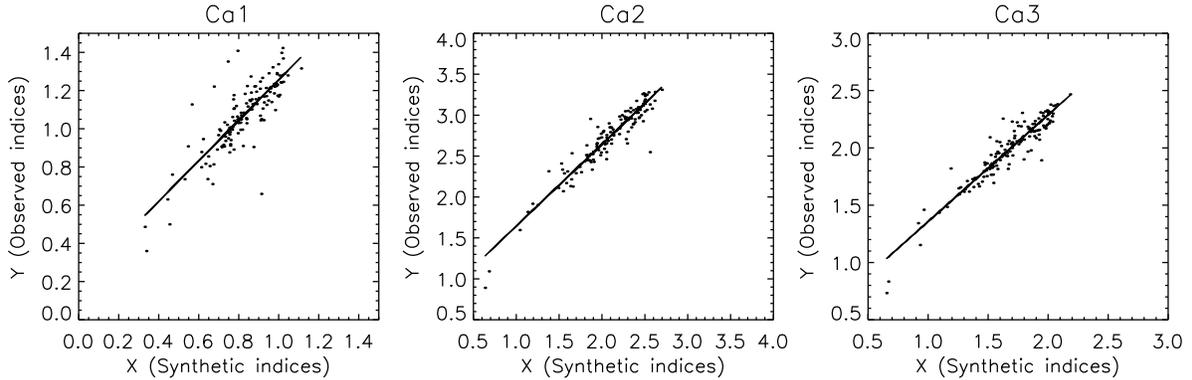}
 \caption[]{Calibration of synthetic NIR Ca{\sc{ii}} triplet indices for dwarfs. This plot shows values of observed indices versus those of the corresponding synthetic indices at the proper terns of atmospheric parameters. The black lines with gradiants close to 1 represent the transformations which could transform our synthetic indices into the observed indices system. The black lines for Ca1, Ca2, and Ca3 is respectively expressed as Y = 1.061X + 0.196, Y = 1.000X + 0.645, and Y = 0.932X + 0.425.}
\end{figure*}

\subsection[]{Comparison between calibrated synthetic and observed INDO-U.S Indices}
In this section, we compare the calibrated synthetic indices with the observational indices measured on the test sample from the INDO-U.S. library. We use $T_{eff}$ as the baseline for comparisons since $T_{eff}$ is the most important and less model-dependent parameter \citep{b3}. The temperature scales of the synthetic and the observational spectra are different which might introduce mismatches between the two scales, so we need to assume a temperature scale $\theta$ = 5040/$T_{eff}$ for both of the synthetic and the observational spectra, and we check the validity of this assumption by studying the behavior of the synthetic and the observational Ca{\sc{ii}} indices. Additionally, we divide the sample into two groups in terms of log $g$. One group is ``dwarfs" with log $g$ $\geqslant$ 3.5 and the other group with log $g$ $<$ 3.5 is ``giants". Then, we perform the comparisons in different [Fe/H] bins seperately for the groups.

For each of the groups, the values of the Ca{\sc{ii}} indices computed on the test sample from the INDO-U.S. library are plotted versus $\theta$ in eight different [Fe/H] bins which have intervals of $\Delta$[Fe/H] = $\pm$0.1 dex respectively centered at -2.5, -2.0, -1.5, -1.0, -0.5, 0.0, +0.2, and +0.5 dex. Then, in each bin, both SSA and NSSA isogravities of the calibrated synthetic indices are overplotted for the sake of comparisons. It should be noted that we revise the interval to be $\Delta$[Fe/H] = $\pm$0.3 and $\pm$0.2 when [Fe/H] centers at -2.5 and 0.5 dex for the group of dwarfs since there would be no suitable stars in these two bins if the interval still maintained to be $\Delta$[Fe/H] = $\pm$0.1 dex. Wheares, for the same reason, the interval is changed to be $\Delta$[Fe/H] = $\pm$0.2 when [Fe/H] centers at -2.5 -1.0, and 0.5 dex for the group of giants. 

Then for dwarfs, Figures 3, 4, 5, and 6 show the spectral index Ca1, Ca2, Ca3, and CaT versus $\theta$ ($\theta$ = 5040/$T_{eff}$) respectively in the eight different metallicity bins. In each of the bins, synthetic SSA isogravities are presented by solid lines and NSSA isogravities are shown as dashed lines while the observational indices from the INDO-U.S. library are presented by black dots. We draw a black dashed line vertically at $T_{eff}$ = 6000 K in every bin to distinguish the indices of hot stars (hotter than 6000 K in the left region to the dashed line; Hot Indices Region) from the indices of cool stars (cooler than 6000 K in the right region to the dashed line; Cool Indices Region). In Hot Indices Region, the observational point positions might not match well with the synthetic isogravities somewhere because of the possible irregularities of the synthetic spectra of SSS which indicates that the synthetic isogravities in Hot Indices Region are perhaps not what they really are. In Cool Indices Region, it is obviously shown that no systematic offset between the synthetic isogravities and the observational point positions exists. This good consistency really confirms that the synthetic and the observational spectra are on the same $\theta$ = 5040/$T_{eff}$ temperature scale and allows us to be confident on the methodology adopted for the comparisons. In general, for the dwarfs cooler than 6000 K, the good agreement of the observational point positions with the synthetic isogravities in each metallicity bin convincingly indicates the correctness of the behavior of our defined synthetic indices versus surface gravity. Furthermore, the good consistencies in all [Fe/H] bins show that the dependency of the observational indices on metallicity could be well reproduced by our synthetic indices. However, a powerful evidence for agreement at very low metallicity ([Fe/H] $<$ -1.0) and/or at effective temperature lower than 4200 K ($\theta$ $>$ 1.2) is unfortunately hampered by few suitable data available from the test sample of the INDO-U.S. library.

As for giants, we similarly plot the spectral index Ca1, Ca2, Ca3, and CaT versus $\theta$ ($\theta$ = 5040/$T_{eff}$) in the eight different metallicity bins overplotted by SSA and NSSA isogravities of the calibrated indices respectively in Figure 7, 8, 9, and 10. In these figures, there seems to be a slightly tendency of the four synthetic indices to be overestimated in all metallicity bins. This less good consistency between the positions of the observational points and the synthetic isogravities most probably stems from the inaccuracy of stellar parameters for the INDO-U.S. giants because the consistency for giants is even much worse if we replace the improved stellar parameters from \citet{b23} by the previous parameters from the INDO-U.S. library itself. Although \cite{b23} has refined the stellar parameters for the INDO-U.S. library, the accuracy is still not enough for giants.  Still, a sound comparison at effective temperature larger than 5000 K and at [Fe/H] $<$ -1.0 and [Fe/H] = 0.5 is hampered by few suitable data from the test sample of the INDO-U.S. library. 
 
On the whole, our calibrated synthetic indices can give a good representation of the observational indices, and such an representation is much better for dwarfs. Simultaneously, we are looking forward to a more precise version of the stellar parameters for the INDO-U.S. library, and particularly for giants. 
\begin{figure*}
\centering
 \includegraphics[scale=3.2]{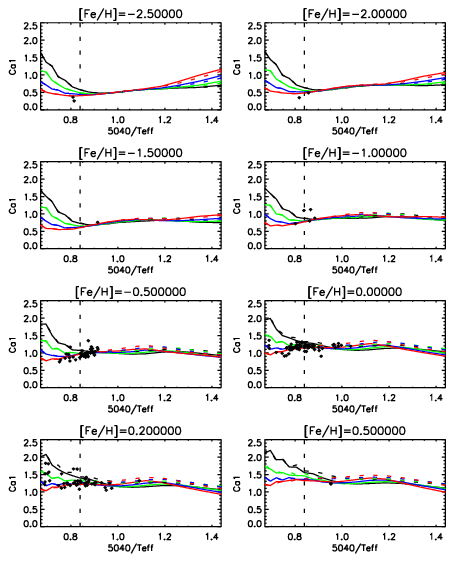}
 \caption[]{Spectral index Ca1 of observed dwarfs stars vs. $\theta$ = 5040/$T_{eff}$ in different metallicity groups. Synthetic SSA (solid lines) and NSSA (dashed lines) isogravities are overplotted. The black, green, blue, and red color represent log $g$ = 3.5, 4.0, 4.5, and 5.0, respectively.}
\end{figure*}

\begin{figure*}
\centering
 \includegraphics[scale=3.2]{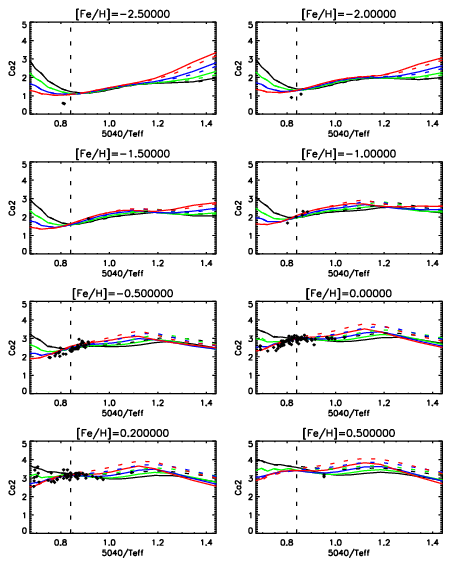}
 \caption[]{Spectral index Ca2 of observed dwarfs stars vs. $\theta$ = 5040/$T_{eff}$ in different metallicity groups. Synthetic SSA (solid lines) and NSSA (dashed lines) isogravities are overplotted. The black, green, blue, and red color represent log $g$ = 3.5, 4.0, 4.5, and 5.0, respectively.}
\end{figure*}

\begin{figure*}
\centering
 \includegraphics[scale=3.2]{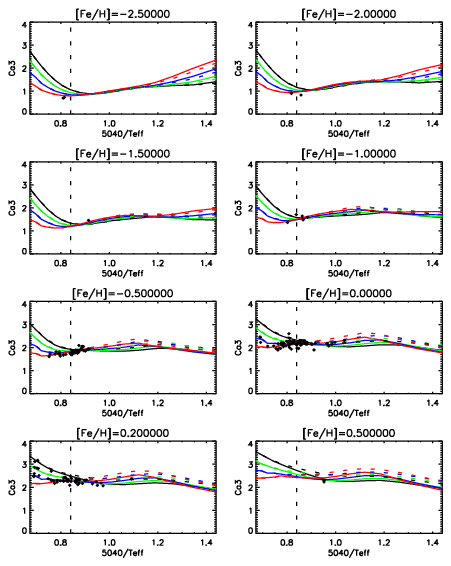}
 \caption[]{Spectral index Ca3 of observed dwarfs stars vs. $\theta$ = 5040/$T_{eff}$ in different metallicity groups. Synthetic SSA (solid lines) and NSSA (dashed lines) isogravities are overplotted. The black, green, blue, and red color represent log $g$ = 3.5, 4.0, 4.5, and 5.0, respectively.}
\end{figure*}

\begin{figure*}
\centering
 \includegraphics[scale=3.2]{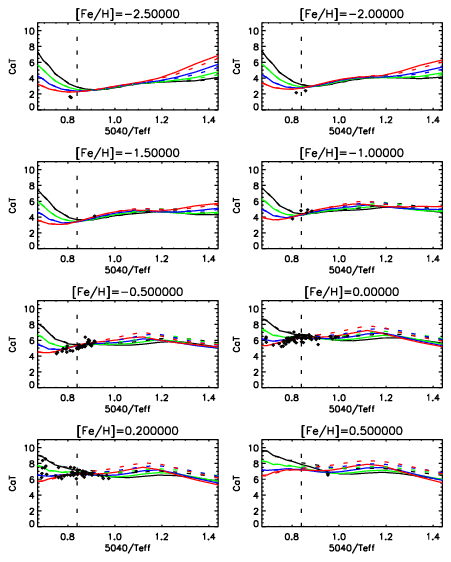}
 \caption[]{Spectral index CaT of observed dwarfs stars vs. $\theta$ = 5040/$T_{eff}$ in different metallicity groups. Synthetic SSA (solid lines) and NSSA (dashed lines) isogravities are overplotted. The black, green, blue, and red color represent log $g$ = 3.5, 4.0, 4.5, and 5.0, respectively.}
\end{figure*}

\begin{figure*}
\centering
 \includegraphics[scale=3.2]{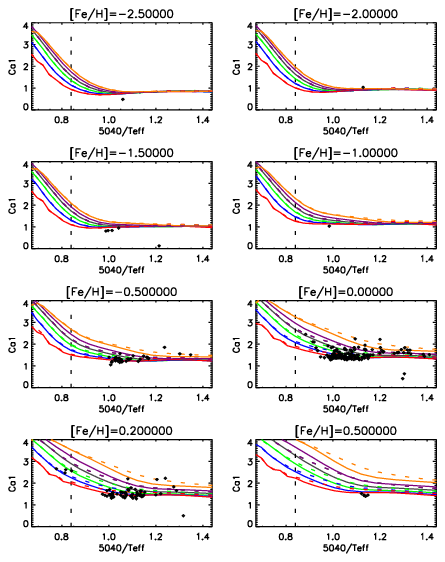}
 \caption[]{Spectral index Ca1 of observed giant stars vs. $\theta$ = 5040/$T_{eff}$ in different metallicity groups. Synthetic SSA (solid lines) and NSSA (dashed lines) isogravities are overplotted. The orange, purple, dark grey, green, blue, and red color represent log $g$ = 0.5, 1.0, 1.5, 2.0, 2.5, and 3.0, respectively.}
\end{figure*}

\begin{figure*}
\centering
 \includegraphics[scale=3.2]{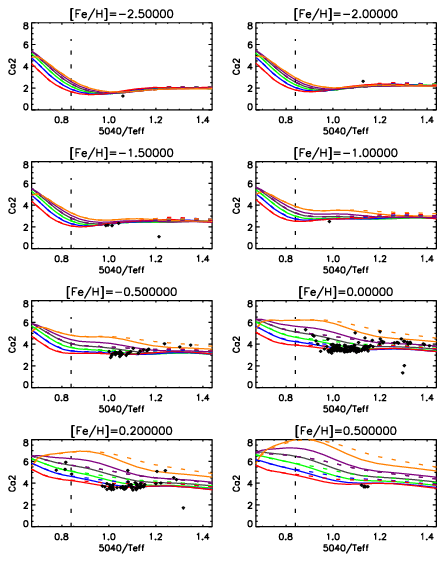}
 \caption[]{Spectral index Ca2 of observed giant stars vs. $\theta$ = 5040/$T_{eff}$ in different metallicity groups. Synthetic SSA (solid lines) and NSSA (dashed lines) isogravities are overplotted. The orange, purple, dark grey, green, blue, and red color represent log $g$ = 0.5, 1.0, 1.5, 2.0, 2.5, and 3.0, respectively.}
\end{figure*}

\begin{figure*}
\centering
 \includegraphics[scale=3.2]{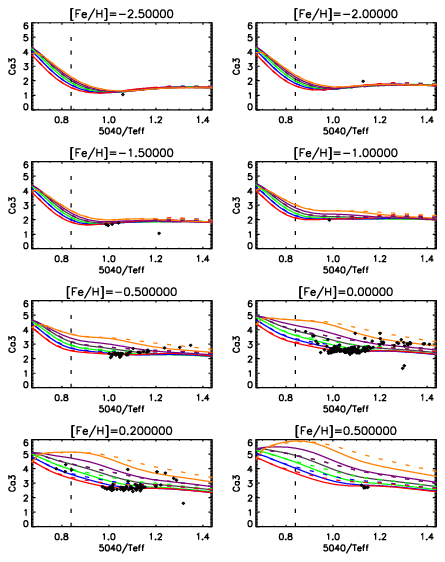}
 \caption[]{Spectral index Ca3 of observed giant stars vs. $\theta$ = 5040/$T_{eff}$ in different metallicity groups. Synthetic SSA (solid lines) and NSSA (dashed lines) isogravities are overplotted. The orange, purple, dark grey, green, blue, and red color represent log $g$ = 0.5, 1.0, 1.5, 2.0, 2.5, and 3.0, respectively.}
\end{figure*}

\begin{figure*}
\centering
 \includegraphics[scale=3.2]{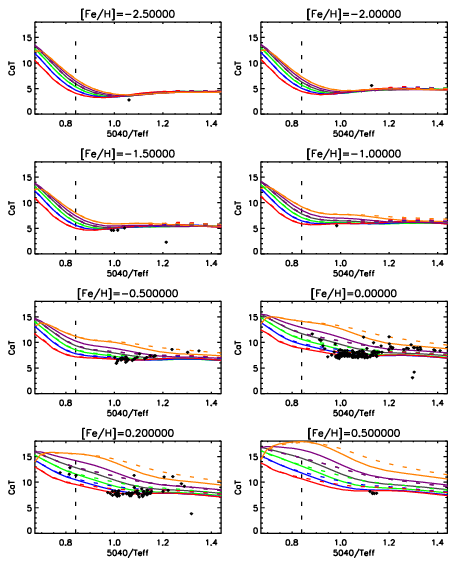}
 \caption[]{Spectral index CaT of observed giant stars vs. $\theta$ = 5040/$T_{eff}$ in different metallicity groups. Synthetic SSA (solid lines) and NSSA (dashed lines) isogravities are overplotted. The orange, purple, dark grey, green, blue, and red color represent log $g$ = 0.5, 1.0, 1.5, 2.0, 2.5, and 3.0, respectively.}
\end{figure*}

\subsection[]{Test the calibrated synthetic indices on the SDSS stars}
In this section, we compare the calibrated synthetic Ca{\sc{ii}} indices with those measured on the SDSS-DR7 and SDSS-DR8 sample. A sample of 11,938 SDSS-DR7 stars (8309 dwarfs and 3629 giants) with atmospheric parameters within our synthetic grid (3500 K $\leqslant$ $T_{eff}$ $\leqslant$ 7500 K, 0.5 dex $\leqslant$ log $g$ $\leqslant$ 5.0 dex and -2.6 dex $\leqslant$ [Fe/H] $\leqslant$ 0.6 dex) is obtained from SDSS-DR7 spectroscopic database. Here, the ``dwarfs" stands for those with surface gravities larger than 3.5 dex while stars with surface gravities smaller than 3.5 dex belong to the ``giants". In addition, another sample of 22,709 spectra of SDSS-DR8 stars (12,497 dwarfs and 10,212 giants) with parameters within the limits of our synthetic grid is also derived from the SDSS-DR8 spectroscopic database.

Although the SNRs of our SDSS sample are all beyond 20, they are still much lower than those of the test sample from the INDO-U.S. library (Details about the effect of SNRs on the NIR Ca{\sc{ii}} indices are presented in Appendix A1). In addition to this, the number of the SDSS sample is quite large. So it is possible and better for us to show the tests by adopting the mean values instead of the individuals since it is always an effective way to diminish the effect of noises by adopting the mean value instead of the individuals under the circumstance that the number of individuals is quite large . Like what have been done in Section 3.1, we fill the eight different metallicity bins with the suitable spectra from our SDSS-DR7 sample, and in the next step, we group stars in each metallicity bin into several narrow bins by the effective temperature ($\theta$ = 5040/$T_{eff}$) parameter at a constant step of $\Delta\theta$ = 0.05. For now, we have gained numbers of sub-bins formed by the limitations of [Fe/H] and $\theta$ together. Then, in each sub-bin where we firstly calculate the mean value of the effective temperatures and indices of all the stars, we could derive a couple of mean values ($Mean_{\theta}$ and $Mean_{index}$) which absolutely represents a cursory generality of overall effective temperatures and indices in this sub-bin. Moreover, the mean values of the SSA and NSSA indices at every $T_{eff}$ point from the synthetic grid are also computed in each of the eight metallicity bins. Then, we draw all couples of the means from all sub-bins in a $Mean_{index}$ v.s. $Mean_{\theta}$ panel, and for the sake of a clear comparison, the means of the SSA and NSSA indices are also overplotted as solid and dashed lines. Such comparisons for dwarfs and giants from the SDSS-DR7 sample are respectively shown in Figures 11 and 12 where every column presents the change of [Fe/H] from -2.5 dex to 0.5 dex and every line exhibits the comparisons respectively for Ca1, Ca2, Ca3, and CaT index. It is obvious that for all of the dwarf cases, the calibrated synthetic indices (solid and dashed lines) can well predict the trends of the indices from the SDSS-DR7 sample with $\theta$ except for those in the bins centering on [Fe/H] = -2.5 and 0.5 dex where there are too few suitable data from the SDSS-DR7 sample. Detailedly, the agreement between synthetic indices and indices from the SDSS-DR7 sample appears to be worse in the bin of [Fe/H]=-2.5$\pm$0.1 dex. As for the bin of [Fe/H]=0.5$\pm$0.1, there are no suitable data from the SDSS-DR7 sample. So we cannot be sure of anything conclusive for indices in these two metallicity bins. In order to solve this problem, we intend to remake the tests on the SDSS-DR8 sample which has more data in the bins of [Fe/H]=-2.5$\pm$0.1 and 0.5$\pm$0.1 dex. Ultimately, the comparisons are shown in Figures 13 and 14 for dwarfs and giants. Evidently, the agreement between the calibrated synthetic and the observatianl SDSS-DR8 indices is also good for all the dwarf cases from the SDSS-DR8 sample but for those in the bins of [Fe/H]=-2.5$\pm$0.1 and 0.5$\pm$0.1 dex where the SDSS-DR8 indices seem to be underestimated. While in the giant cases (Figures 12 and 14), we fail to see a good agreement between the synthetic and the SDSS indices because all of the SDSS indices fall below the theoretical lines which still most probably originate from the inaccuracy of the stellar parameters generated by the SSPP pipeline for SDSS stars. 

Conlusively, it is good to use SDSS dwarfs to check the validity of our synthetic indices; however, it is a little bad when SDSS giants are used. The reason is most likely to be the inaccuracies of the stellar parameters generated by the SSPP pipeline for the SDSS-DR7 and SDSS-DR8 giants. So we hope for more precise stellar parameters for SDSS stars, especially for giants. Simultaneously, we are looking forward to the forthcoming LAMOST spectra which are expected to have more accurate stellar parameter estimations through a completely new and improved 2-D and 1-D stellar parametrization pipelines. 

\begin{figure*}
\centering
 \includegraphics[scale=2.2]{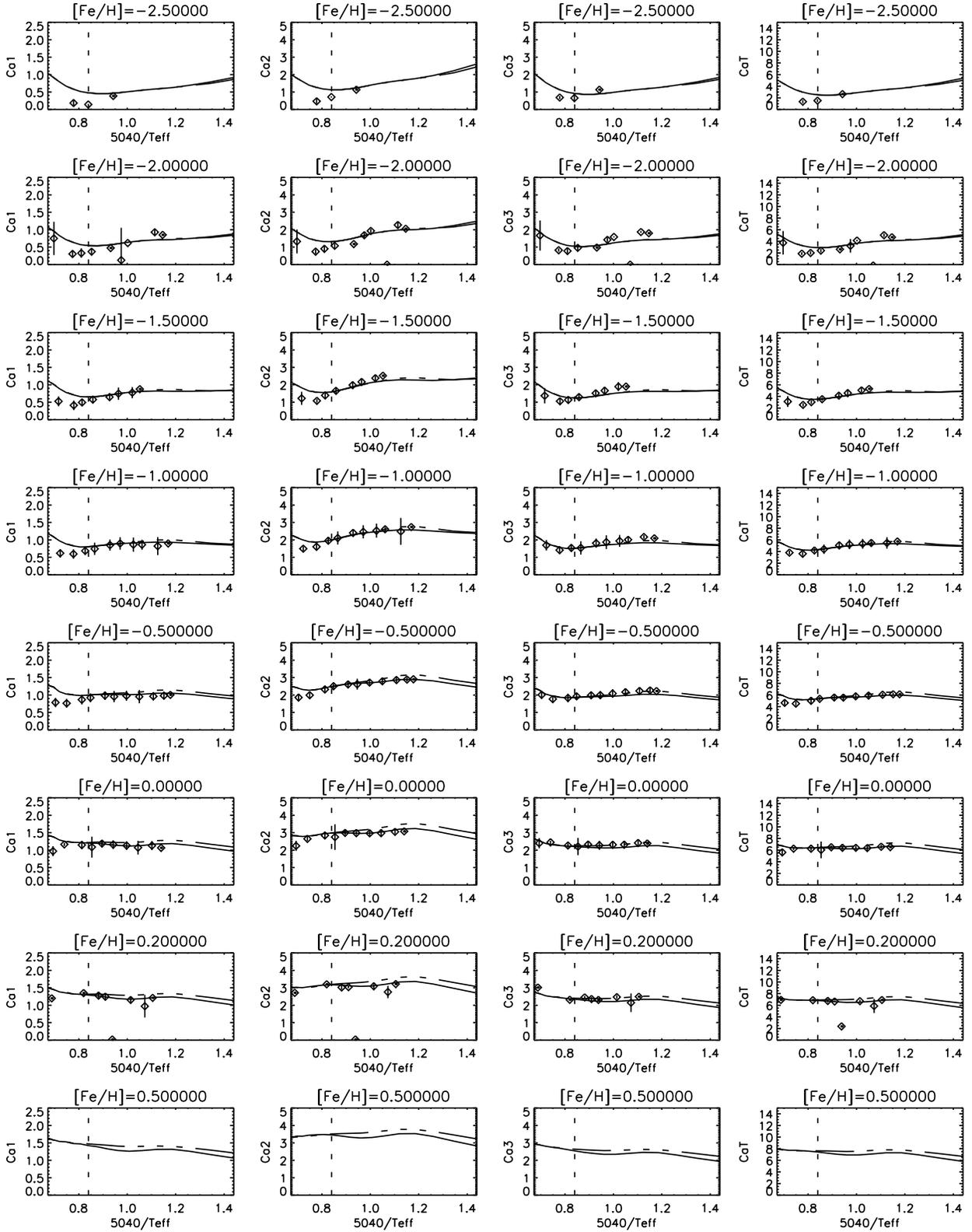}
 \caption[]{Mean values of spectral index Ca1, Ca2, Ca3 and CaT of SDSS-DR7 dwarfs v.s. $\theta$=5040/$T_{eff}$ in different metallicity groups (columns). Mean synthetic SSA (solid lines) and NSSA (dashed lines) are overplotted for the clarity of agreement between our calibrated synthetic indices and the observed SDSS-DR7 ones.}
\end{figure*}
\begin{figure*}
\centering
 \includegraphics[scale=2.2]{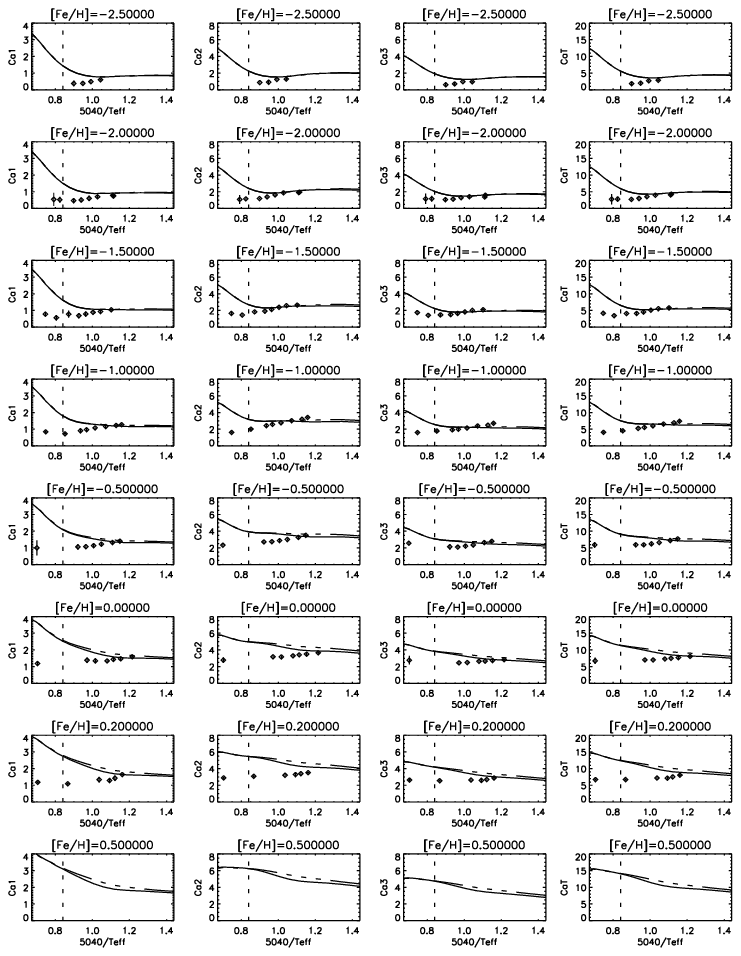}
 \caption[]{Mean values of spectral index Ca1, Ca2, Ca3 and CaT of SDSS-DR7 giants v.s. $\theta$=5040/$T_{eff}$ in different metallicity groups (columns). Mean synthetic SSA (solid lines) and NSSA (dashed lines) are overplotted for the clarity of agreement between our calibrated synthetic indices and the observed SDSS-DR7 ones.}
\end{figure*}
\begin{figure*}
\centering
 \includegraphics[scale=2.2]{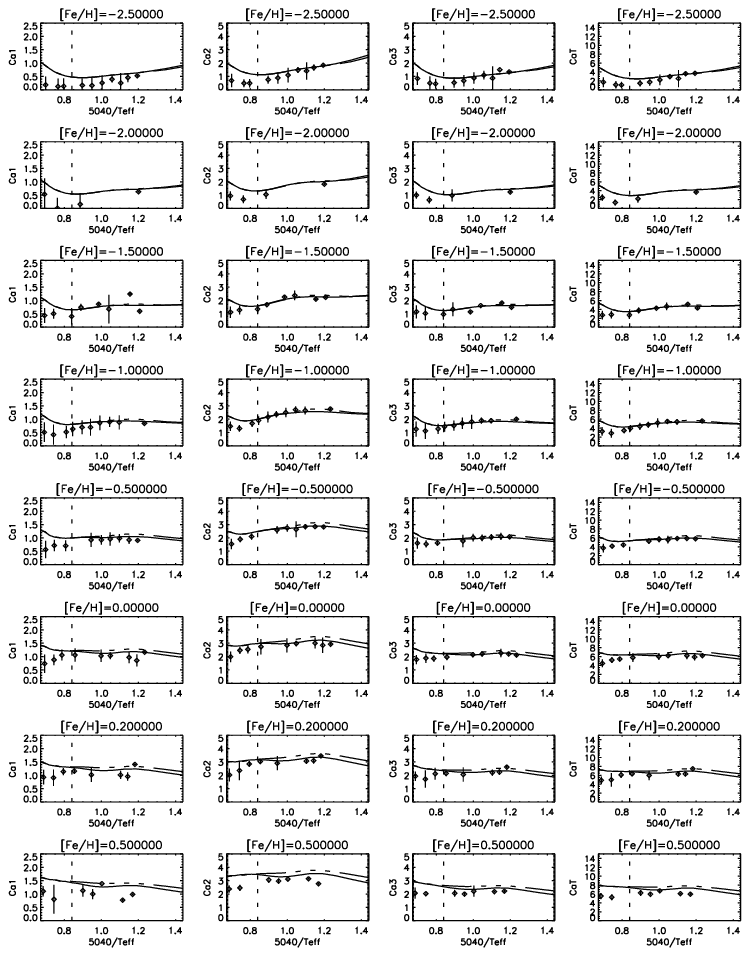}
 \caption[]{Mean values of spectral index Ca1, Ca2, Ca3 and CaT of SDSS-DR8 dwarfs v.s. $\theta$=5040/$T_{eff}$ in different metallicity groups (columns). Mean synthetic SSA (solid lines) and NSSA (dashed lines) are overplotted for the clarity of agreement between our calibrated synthetic indices and the observed SDSS-DR8 ones.}
\end{figure*}
\begin{figure*}
\centering
 \includegraphics[scale=2.2]{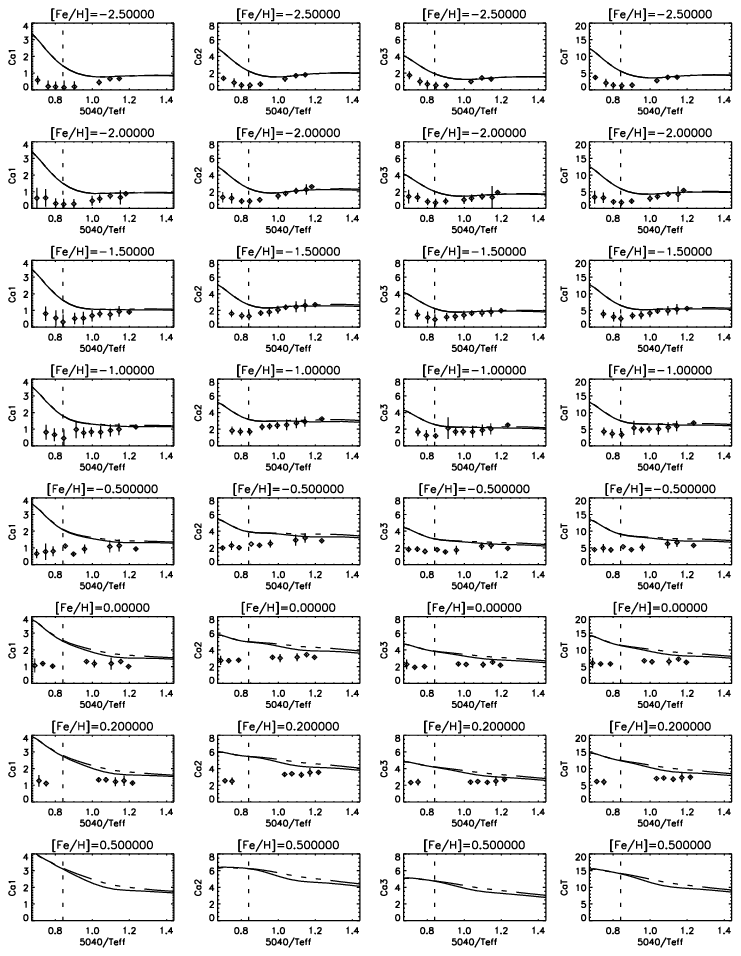}
 \caption[]{Mean values of spectral index Ca1, Ca2, Ca3 and CaT of SDSS-DR8 giants v.s. $\theta$=5040/$T_{eff}$ in different metallicity groups (columns). Mean synthetic SSA (solid lines) and NSSA (dashed lines) are overplotted for the clarity of agreement between our calibrated synthetic indices and the observed SDSS-DR8 ones.}
\end{figure*}

\section[]{Relations between Ca{\sc{ii}} indices and stellar atmospheric parameters}
\subsection[]{Relations with surface gravity}
We investigate how dependent each of the indices (Ca1, Ca2, Ca3, and CaT) is on the parameter of log $g$ itself, and exhibit the investigation results in Figure 15 where the strength of an index is plotted versus log $g$ for both the metal-poor ([Fe/H] $<$ 0.0) and the metal-rich ([Fe/H] $\geqslant$ 0.0). Apparently, for dwarfs (log $g$ $\geqslant$ 3.5; to the right region of the dashed line in Figure 15), there is little effect of log $g$ on the four indices no matter whether the star is poor or rich in metallicity. Therefore, we could infer that, for dwarfs, the scatters of the indices in Figure 15 should be probably led by the effect of solely [Fe/H] or $T_{eff}$, or the combination of [Fe/H] and $T_{eff}$. Then, for giants (log $g$ $<$ 3.5; to the left region of the dashed line in Figure 15), indices strengths as a whole appear to be weakened by the increase of log $g$, which stands out for a general but negative influence of log $g$ on index strength. Influence of such a kind looks more serious for the metal-rich than the metal-poor. Therefore, for giants also, the scatters of the indices at a same log $g$ value should originate from the effect of solely [Fe/H] or $T_{eff}$, or the combination of [Fe/H] and $T_{eff}$. 

Furthermore, we find that indices strengths as a whole seem to be strengthened by the enhancement of [Fe/H] because the general level of the distributions of indices strengths of the metal-rich (the average line in the right pannel in Figure 15) is obviously higher than that of the metal-poor (the average line the left pannel in Figure 15), comparing the former with the latter. 

In short, it is revealed by Figure 15 that, for dwarfs, log $g$ has a too little effect to be neglected on the indices strengths. Instead, [Fe/H] and $T_{eff}$ play the dominant role; for giants, log $g$ starts to affect indices strengths negatively which means that if log $g$ increased, the indices strengths would decrease. As for [Fe/H], in general, the enhancement of  metallicity can lead the indices strengths to be strengthened for both ``dwarfs" and ``giants". 
\begin{figure*}
\centering
 \includegraphics[scale=2.6]{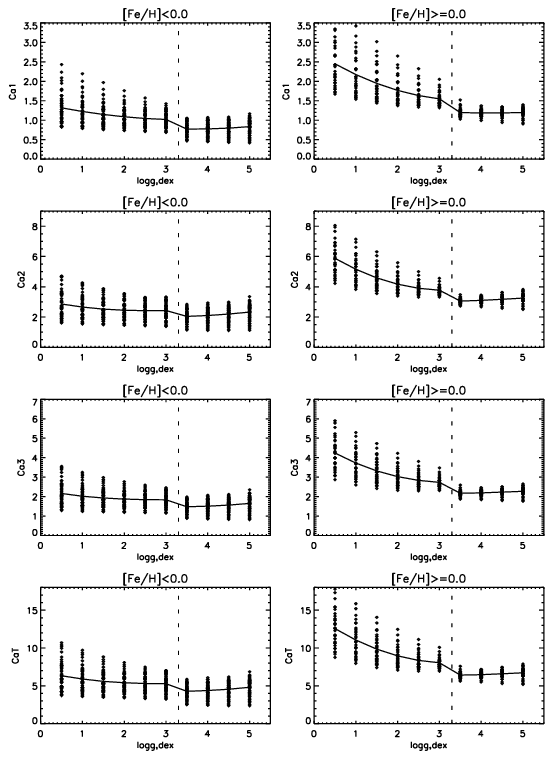}
 \caption[]{Dependence of the Ca1, Ca2, Ca3, and CaT index on the log $g$ parameter. The calibrated LSS SSA indices are represented by the black points and the smoothed line represents the mean value which could give the general direction of the changes. The dashed line is a divide for dwarfs and giants. }
\end{figure*}
\subsection[]{Relations with metallicity}
We show the dependencies of Ca1, Ca2, Ca3, and CaT index on the parameter of [Fe/H] in Figure 16 where the strength of each of the indices is plotted versus [Fe/H], respectively, for dwarfs and giants. Obviously, the four indices are being strengthened by the enhancement of [Fe/H] for both dwarfs and giants. However, for giants, in accord with Figure 16, the scatters of indices stengths at a fixed [Fe/H] appear more serious, and should probably stem from the effect of solely log $g$ or $T_{eff}$, or the combination of the two parameters.

For dwarfs, considering the conlusion of log $g$ having little effect on the four indices strengths from Figure 15, and then combining with Figure 16, we could easily infer that it is [Fe/H] and $T_{eff}$ which affect the indices strengths together while log $g$ plays an ignorable role. 

Then, for giants, [Fe/H], $T_{eff}$, and log $g$ influence the indices together in strength. To be more detailed, for metal-rich giants, log $g$ has a more serious effect on the indices strength than for the metal-poor giants, convinced by the much larger scatters of metal-rich giant indices stengths than those of the metal-poor giants in Figure 16.
\begin{figure*}
\centering
 \includegraphics[scale=2.6]{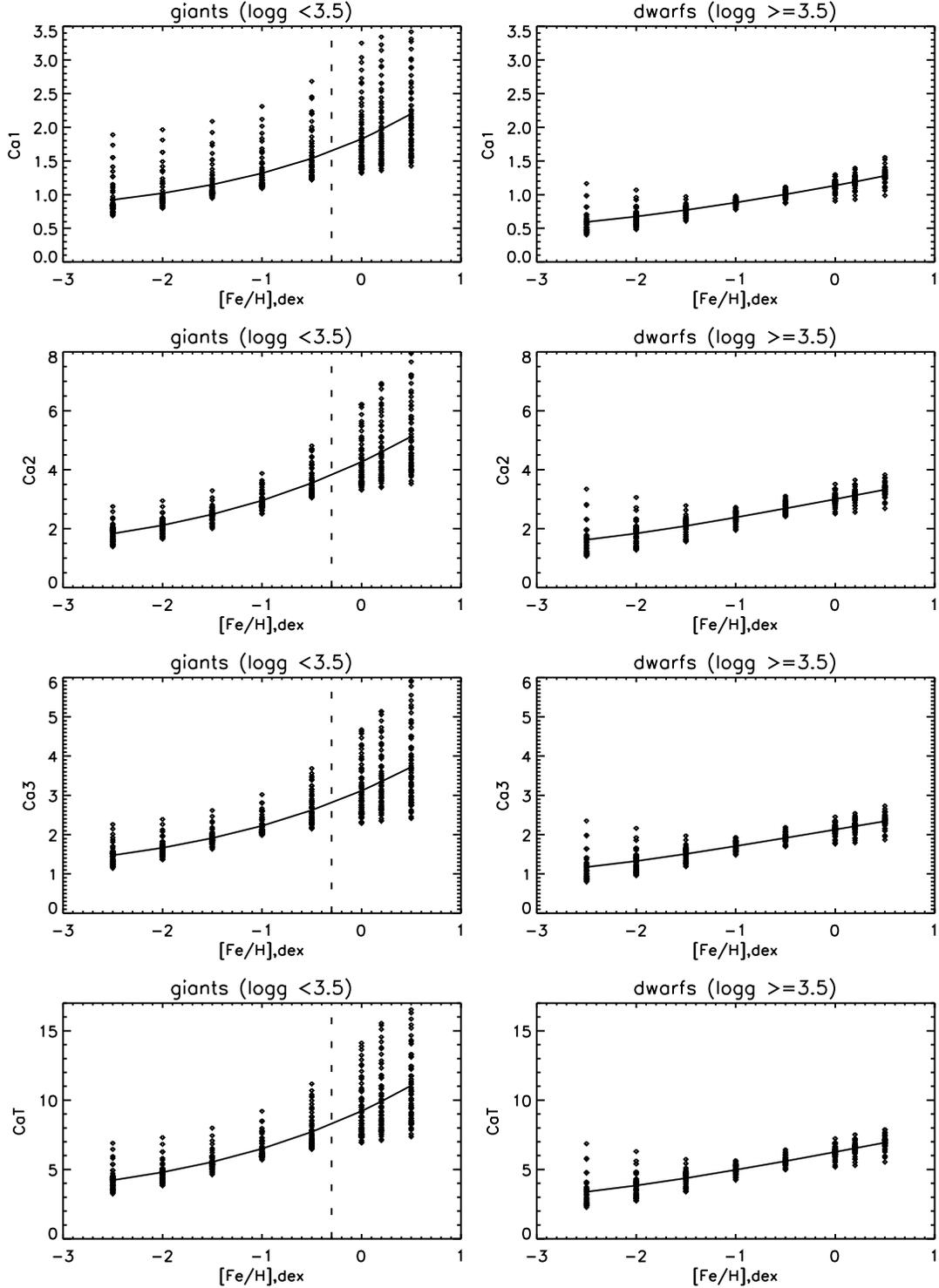}
 \caption[]{Dependence of the Ca1, Ca2, Ca3, and CaT index on the [Fe/H] parameter. The calibrated LSS SSA indices are represented by the black points and the smooth line represents the mean value which could give the general trend of variations. The vertical dashed line is a divide of metal-poor and metal-rich stars. }
\end{figure*}
\subsection[]{Relation with effective temperature}
We do try harder to find out how the parameter of $T_{eff}$ affects the NIR Ca{\sc{ii}} indices via investigating how the indices stengths of SSA spectra change when $T_{eff}$ changes. First, we divide the SSA spectra into four groups ( metal-rich dwarfs, metal-poor dwarfs, metal-rich giants, and metal-poor giants), and then measure the four indices on all groups. For each of the four indices, we calculate the average of index strengths  over $\theta$ ($\theta$ = 5040/$T_{eff}$) for each of the four groups, which are all depicted in Figure 17. Obviously, $T_{eff}$ indeed has an effect on all indices, and only for the metal-rich giants presented by lines crossed with the `plus' symbols in Figure 17, the effect appears obviously. 

To be summarized, the effect of the parameter of $\theta$ ($\theta$ = 5040/$T_{eff}$) on Ca{\sc{ii}} indices is obvious for the metal-rich giants. For the sake of a clear but quanlitative illustration, the impact of the parameters of log $g$ and [Fe/H] on the four indices is generalized in Table 2  where the `little', `much', or `more' means that the parameter has little , much, or more effect on the four indices. The sign of `+' or `-' respectively stands for a  positive or negative trend of the effect of the parameter on the four indices. 

\begin{table}
 \begin{minipage}{120mm}
  \caption{The qualitative effect of stellar parameters on Ca{\sc{ii}} indices. }
  \begin{tabular*}{120mm}{@{}lcccc}
\hline
 Para. &\multicolumn{2}{c}{Dwarfs} &\multicolumn{2}{c}{Giants} \\
       &Metal-rich &Metal-poor       &Metal-rich &Metal-poor       \\
 
\hline
log $g$     &little      &little     &more (-)  &much (-)  \\
$[Fe/H]$    &much (+)    &much (+)   &more (+)  &more (+)  \\
\hline   
\end{tabular*}
\end{minipage}
\end{table}

\begin{figure*}
 \centering
 \includegraphics[scale=2.2]{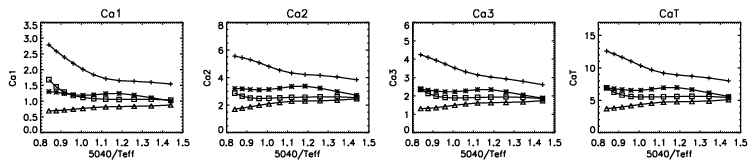}
 \caption[]{Indices of the Ca1, Ca2, Ca3, and CaT versus $\theta$ = 5040/$T_{eff}$. Lines crossed with symbols of star, triangle, plus, and square respectively represent the mean indices values of the metal-rich dwarfs ([Fe/H] $\geqslant$ 0.0, log $g$ $\geqslant$ 3.5), the metal-poor dwarfs ([Fe/H] $<$ 0.0, log $g$ $\geqslant$ 3.5), the metal-rich giants ([Fe/H] $\geqslant$ 0.0, log $g$ $<$ 3.5), and the metal-poor giants ([Fe/H] $<$ 0.0, log $g$ $<$ 3.5).}
\end{figure*}

\subsection[]{Feature strength as a function of stellar atmospheric paramters}
In this section, we regress the relations of the index strength of CaT with stellar atmospheric parameters for synthetic stars, and then, test the derived relations on the observational stars respectively from INDO-U.S. library,  SDSS-DR7, and SDSS-DR8 spectroscopic database which have already been used in the previous part of this paper. The regressions are made by the REGRESS program in IDL software. Suggested by the conclusions from Figures 15 $\sim$ 17, for dwarfs, we expect to express the index strength of CaT by a function of metallicity ([Fe/H]) and effective temperature ($\theta$), and for giants, by a function of all the metallicity ([Fe/H]), surface gravity (log $g$), and effective temperature ($\theta$). The functions are below:  

For dwarfs (log $g$ $\geqslant$ 3.5)

\begin{mathletters}

EW(CaT) = 3.648($\pm$0.134)[Fe/H] - 1.081($\pm$0.157)$\theta$ - 2.235($\pm$0.119)[Fe/H]$\theta$ + 7.460 

\end{mathletters}

For metal-poor giants (log $g$ $<$ 3.5 and [Fe/H] $<$ 0)

\begin{mathletters}
EW(CaT)=4.550($\pm$0.211)[Fe/H] - 2.920($\pm$0.176)log $g$ - 6.765($\pm$0.392)$\theta$ + 1.769($\pm$0.147)log $g$ $\theta$ - 1.966($\pm$0.178)[Fe/H]$\theta$ - 0.354($\pm$0.037)log $g$ [Fe/H] + 17.450
\end{mathletters}

For metal-rich giants (log $g$ $<$ 3.5 and [Fe/H] $\geqslant$ 0)

\begin{mathletters}
EW(CaT)=11.955($\pm$1.467)[Fe/H] - 5.491($\pm$0.341)log $g$ - 12.227($\pm$0.646)$\theta$ + 3.603($\pm$0.297)log $g$ $\theta$ - 5.304($\pm$1.234)[Fe/H]$\theta$ - 1.445($\pm$0.258)log $g$ [Fe/H] + 25.212
\end{mathletters}

The errors in the brackets are the 1-sigma uncertainty estimates for the corresponding coefficients. The coefficients and the corresponding 1-sigma errors are directly returned from the REGRESS program.  

Using the functions above, we predict the strengths of CaT index on dwarfs, metal-poor giants, and metal-rich giants respectively from the previously used SSA, INDO-U.S., SDSS-DR7, and SDSS-DR8 spectral sample. Then we compare predictions with the index strengths directly from measurements. For dwarfs, metal-poor giants, and metal-rich giants, the comparisons are respectively shown in Figures 18, 19, and 20 where the `sigma' in every panel presents the standard deviation of differences between predictions and measurments.

Previously in Section 3, the agreement between the observed Ca{\sc{ii}} indices and our synthetic ones is not so good for giants as that for dwarfs, which inevitably causes the different precisions of the functions above. From Figure 18, the analytical relation of the strength of the CaT index with the stellar parameters for dwarfs is good for both the synthetic SSA dwarfs and the INDO-U.S. dwarfs with small 1-sigma errors. As for SDSS stars, the 1-sigma error turns a little larger for DR7 and DR8 stars because we could not obtain the more accurate parameters for SDSS stars so that the parameters we have to use are from SSPP which are only estimations with a low accuracy. We should have a much better consistency between predictions and measurements for SDSS stars if more accurate parameters could be obtained. The distribution has a up-shift from the gradient 1 line for SDSS-DR8 indices which might be caused by a systematic difference of the different pipeline for data reduction from that for SDSS-DR7 spectra. In Figure 19 for the metal-poor giants, the function is still good for the SSA spectra while it still works for the INDO-U.S., SDSS-DR7, and SDSS-DR8 spectra despite a small deviation when the strength of the CaT index is smaller than 5. In Figure 20 for the metal-rich giants, the function seems good for the SSA, INDO-U.S., and even SDSS-DR7 sample while it does not work well for the SDSS-DR8 ones with the strength of the CaT index smaller than 5. Generally speaking, our functions of estimating the strength of the CaT index with the stellar parameters works for dwarfs, metal-poor giants, and metal-rich giants of late-types with parameters (3500 $\leqslant$ $T_{eff}$ $\leqslant$ 7500 K, -2.5 $\leqslant$ [Fe/H] $\leqslant$ 0.5 and 0.5 $\leqslant$ log $g$ $\leqslant$ 5.0) as accurate as possible. These functions are simple and approximately good since the 1-sigma errors can be accepted.    \\





\begin{figure*}
 \centering
 \includegraphics[scale=2.2]{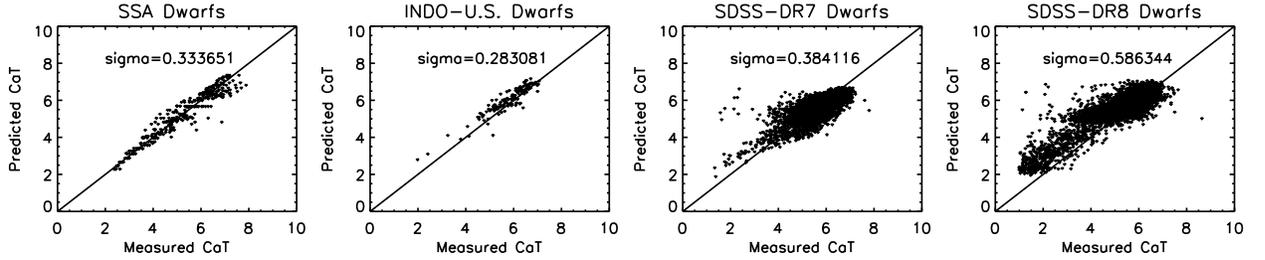}
 \caption[]{Predicted versus measured EW(CaT) for dwarfs. The predicted values have been calculated as functions of atmospheric parameters. The straight line has a gradient of 1. }
\end{figure*}
\begin{figure*}
 \centering
 \includegraphics[scale=2.2]{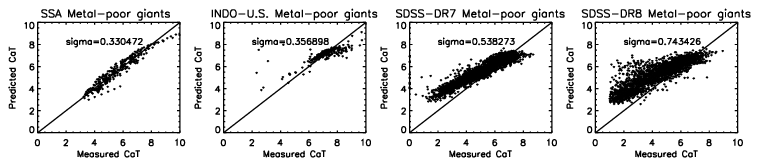}
 \caption[]{Predicted versus measured EW(CaT) for metal-poor giants. The predicted values have been calculated as functions of atmospheric parameters. The straight line has a gradient of 1. }
\end{figure*}
\begin{figure*}
 \centering
 \includegraphics[scale=2.2]{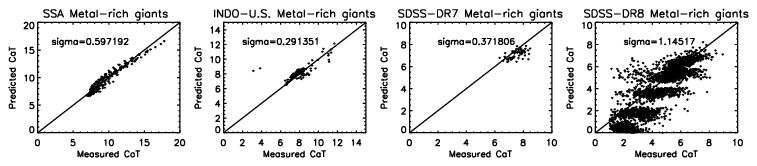}
 \caption[]{Predicted versus measured EW(CaT) for metal-rich giants. The predicted values have been calculated as functions of atmospheric parameters. The straight line has a gradient of 1. }
\end{figure*}

\section[]{Comparisons with the CaHK index}
The CaHK index originating from the Ca{\sc{ii}} Fraunhofer H and K features (Ca H($\lambda$ 3936) and K($\lambda$ 3934)) can also be used to measure calcium abundance in galaxies. This index has been defined in \citet{b24} and \citet{b3}, respectively. The Ca H and K doublet features are strong in the integrated-light spectra of cool stars and elliptical galaxies largely composed of low-mass and cooler stars. 

Comparing with Ca{\sc{ii}} triplet, Ca H and K give much stronger percentage response to increased Ca abundance, and the response spreads over a much broader wavelength region which could make measurements even easier (details in Figure 1 in \citet{b25}). Furthermore, it is noted that the CaHK index is a factor of three more sensitive than the Ca{\sc{ii}} triplet indices \citep{b24}. Unlike the Ca{\sc{ii}} triplet indices, the CaHK index is free of the contamination by the Paschen lines and the NLTE impact (details in Sections 2.1 and 2.2). However, the soundness of using the CaHK index as an ideal calcium measurement tool is hampered by the emission components often present within the centers of the H and K absorption lines for stars of type G0 and later \citep{b26}. In specific, \citet{b27} has furthermore found out the linear correlation of the logarithms of the emission-line widths with the spectroscopic absolute magnitudes. This Wilson-Bappu effect surely leads to weaker measurements of Ca H and K equivalent width but has no effects on measurements of Ca{\sc{ii}} triplet widths. Additionally, the Ca{\sc{ii}} triplet features appear around 8600 $\AA$ in the NIR wavelength band where the absorbing effects of dust are small; instead, the Ca H and K doublet features occur around 3900 $\AA$ in the very blue wavelength band where indices are more seriously subject to absorption by the interstellar medium.

In \citet{b3}, it is shown that the CaHK index is sensitive to $\alpha$-element enhancement which influences the CaHK index by strengthening it (Figures 1 in \citet{b3}). In our work, we have verified that the Ca{\sc{ii}} triplet indices are almost independent on $\alpha$-element enhancement (Figure 1). For dwarfs, the behaviors of Ca{\sc{ii}} triplet indices with stellar parameters have a little better reproduced those of observational ones as can be deduced by the more compact consistency with the isogravities in Figures 3 $\sim$ 6 while the behaviors of the CaHK index are not so good as observed (Figure 6 in \citet{b3}). For giants, the Ca{\sc{ii}} triplet indices perform still a little better behaviors with stellar parameters (Figures 7$\sim$ 10) than the CaHK index (Figure 7 in \citet{b3}). Moreover, the light of the Ca H and K features is mostly contributed by main-sequence and warmer giants, in contrast to the light of the Ca{\sc{ii}} triplet features, which is dominated by red giant branch tip and asymptotic giant branch \citep{b25}. 

Briefly, the Ca H and K doublet are important features in the blue wavelength band, and so are the Ca{\sc{ii}} triplet in the NIR wavelength band. Both of them can be used together in spectra where both of the CaHK and the Ca{\sc{ii}} index are prominent in order to get a more convincing and consistent conclusion; they can be both useful tools to measure calcium abundance in galaxies under different circumstances and for different aims; however, once any of the two indices is adopted, it should be better to take its advantages and disadvantages into considerations.

\section[]{Summary}
In this work, we define a new set of NIR Ca{\sc{ii}} triplet indices and a new index CaT as the sum strength of the triplet lines. Then we measure every index of the four on the synthetic spectra, with the same spectral resolution (R = 1800) and wavelength scale (approximately 2.0) as SDSS and the forthcoming LAMOST spectra, which are synthesized by SPECTRUM, a spectral synthetic program, using the ATLAS9 model grid of which the effective temperature ($T_{eff}$) parameter ranges from 3500 to 7500 K with a step of 250 K, and the surface gravity (log $g$) covers from 0.5 to 5.0 dex with an interval of 0.5 dex. As for metallicity, two assumptions are adopted to model the chemical composition, Solar Scaled Abundances (SSA) with [$\alpha$/Fe] = 0 and Non-Solar Scaled Abundances (NSSA) with enhanced $\alpha$ element ([$\alpha$/Fe] = 0.4 dex). In the SSA model, 8 values are adopted for metallicity ([Fe/H] = -2.5, -2.0, -1.5, -1.0, -0.5, 0.0, 0.2, 0.5) while in the NSSA model, a poorer value of [Fe/H] = -4.0 dex is supplemented. After measurement, we compare the Ca{\sc{ii}} triplet indices from the SSA model with those from the NSSA one at the same stellar parameters. Such a comparison indicates that the Ca{\sc{ii}} triplet indices are all quasi-independent on $\alpha$ element. Later on, we calibrate our synthetic Ca{\sc{ii}} triplet indices with the observed stars of the INDO-U.S. library. The most important part of this work is the comparisons of our calibrated synthetic indices with the observational ones which can be summarized as follows.

1. We make this comparison between our calibrated synthetic Ca{\sc{ii}} indices and the observational ones of INDO-U.S. and SDSS DR7 and DR8 stars with the parameters within the synthetic grid. Comparison is respectively made for dwarf and giant stars. 

2. For dwarfs, the behaviours of our calibrated synthetic indices versus effective temperature, surface gravity and metallicity well reproduce that of the observational indices of INDO-U.S., SDSS DR7 and DR8 stars. It is worthwhile to stress that this good agreement is due to the fact that both theoretical and observational indices are on the same temperature scale.

3. For giants, the agreement between the synthetic indices and the observational indices is not so good as that for dwarfs. This discrepancy might be the result of our inaccurate calibration caused by a little larger uncertainty of stellar parameters for giants. 

Therefore, we present a synthetic library of the NIR Ca{\sc{ii}} triplet indices and a summed index CaT which is quite appropriate and more trustful for fully exploiting the near-infrared content of the SDSS and the forthcoming LAMOST stellar spectroscopic database which represents and will represent the most extended collection of stellar spectra up to now.   

The relations of the four Ca{\sc{ii}} indices with stellar atmospheric parameters (log $g$, [Fe/H] and $T_{eff}$) have also been investigated qualitatively in the second half of this paper. Firstly, surface gravity (log $g$) has little effect on Ca{\sc{ii}} indices of dwarfs, but much effect of a negetive trend on those of metal-poor giants and even more on those of metal-rich giants. Secondly, metallicity ([Fe/H]) affects much on the Ca{\sc{ii}} indices of dwarfs and more on those of giants by a positive trend. Speaking of effective temperature ($T_{eff}$), $\theta$ ($\theta$ = 5040/$T_{eff}$) certainly has an effect on the Ca{\sc{ii}} indices of both dwarfs and giants but by an irregular trend. Ultimately, we derive an function to approximately estimate the strength of the CaT index. The function for dwarfs well expresses the relation of the CaT index with the $T_{eff}$ and [Fe/H] parameters and is tested on the SSA, INDO-U.S., SDSS-DR7, and SDSS-DR8 dwarfs. The predictions by the function are generally consistent with the measurements within the sigma errors which we could be accepted. All of the functions are applying for stars with more accurate stellar parameter estimates. Absolutely, the gain of the more accurate parameters for stars particularlly for giants from the SDSS database is the key to improve our test of Ca{\sc{ii}} indices on the SDSS giants. Therefore, a new method to improve the precision of stellar parameters for the SDSS stars is expected. Simultaneously, we are looking forward to a large number of LAMOST stellar spectra with refined parameters to check the validity of our synthetic library of the Ca{\sc{ii}} indices on LAMOST spectra. 

In addition, we compare the differences between the Ca{\sc{ii}} triplet and the Ca H and K doublet indices in several major aspects.   

Finally, as a supplement of this paper, we investigate the effect of SNR to the new set of Ca{\sc{ii}} indices in the Appendix, which shows that the impact of SNR on the indices is so weak that we can ignore the effect when SNRs are beyond 20.

\acknowledgments

We would like to thank the referee for very useful comments and suggestions and thank Fengfei Wang of NAOC for very helpful discussions. 
This study is supported by the Natural Science Foundation of China (NSFC) under No. 10973021 grants.





\appendix

\section[]{Sensitivity of the Ca{\sc{ii}} triplet indices to SNR}
The database on which we define our NIR Ca{\sc{ii}} triplet indices is made up of the synthetic spectra which have no noises. In this Appendix, we discuss the sensitivity of Ca1, Ca2, Ca3, and CaT index to the spectral SNR. 
First we regard the SNR between 8620 and 8630 $\AA$ of a synthetic spectrum as the SNR of the whole synthetic spectrum because the continuum is fairly smooth and few features exist in that wavelength region where the flux level of the continuum is quite close to one since our synthetic spectra are already normalized. In a clean theoretical spectrum, the flux of a pixel is the pure signal of the pixel. So the strength of the signal of the smooth continuum in a normalized theoretical spectrum is undoubtfully approximate to one. Then we can derive any SNR value we expect by adding random normally distributed Gaussian noises with standard deviations equal to the corresponding values to the whole synthetic spectrum. For example, if we add a random Gaussian noise with the mean and standard deviation respectively equal to 1 and 0.2, we can obtain a noise-added synthetic spectrum with SNR = 1/0.2 =5. If the standard deviation is chosen to be 0.05, the SNR goes to be 20 (SNR = 1/0.05). In this test, we derive 23 noise-added synthetic spectrum with different SNR (SNR = 1, 3, 5, 10, 15, 20, 25, 30, ..., 95, 100) by adding normally-distributed Gaussian noises with the distinguished standard deviation values to one synthetic spectrum. We compute all of the four indices on each of the 23 different-SNR synthetic spectrum (Ca$*$) and those on the original synthetic spectrum with no noise added to (Ca). We do the same for every synthetic spectrum of our LSS SSA. Finally, Ca$*$ and Ca are compared for all the SNRs. The comparison shows that the four indices are insensitive to SNR if SNRs are larger than 25, a little sensitive if SNRs are between 15 and 25 and very sensitive if SNRs fall below 15. However, CaT index is the sum of the Ca{\sc{ii}} triplet indices which means noises are summed simultaneously. So it should be more sensitive to SNRs. It seems that CaT index is affected more by SNR when SNRs are below 40. For NSSA, we get the same conclusions. Therefore, the lower the SNR is, and the less trust the four indices deserve. Finally, Figures A1 $\sim$ A4 are presented to respectively show the changes of sensitivities of Ca1, Ca2, Ca3 and CaT to SNRs. In every figure, we just show comparisons when SNRs are 15, 20, 25, 30, 40, 60, 80 and 100.

\begin{figure*}
\centering
 \includegraphics[scale=1.8]{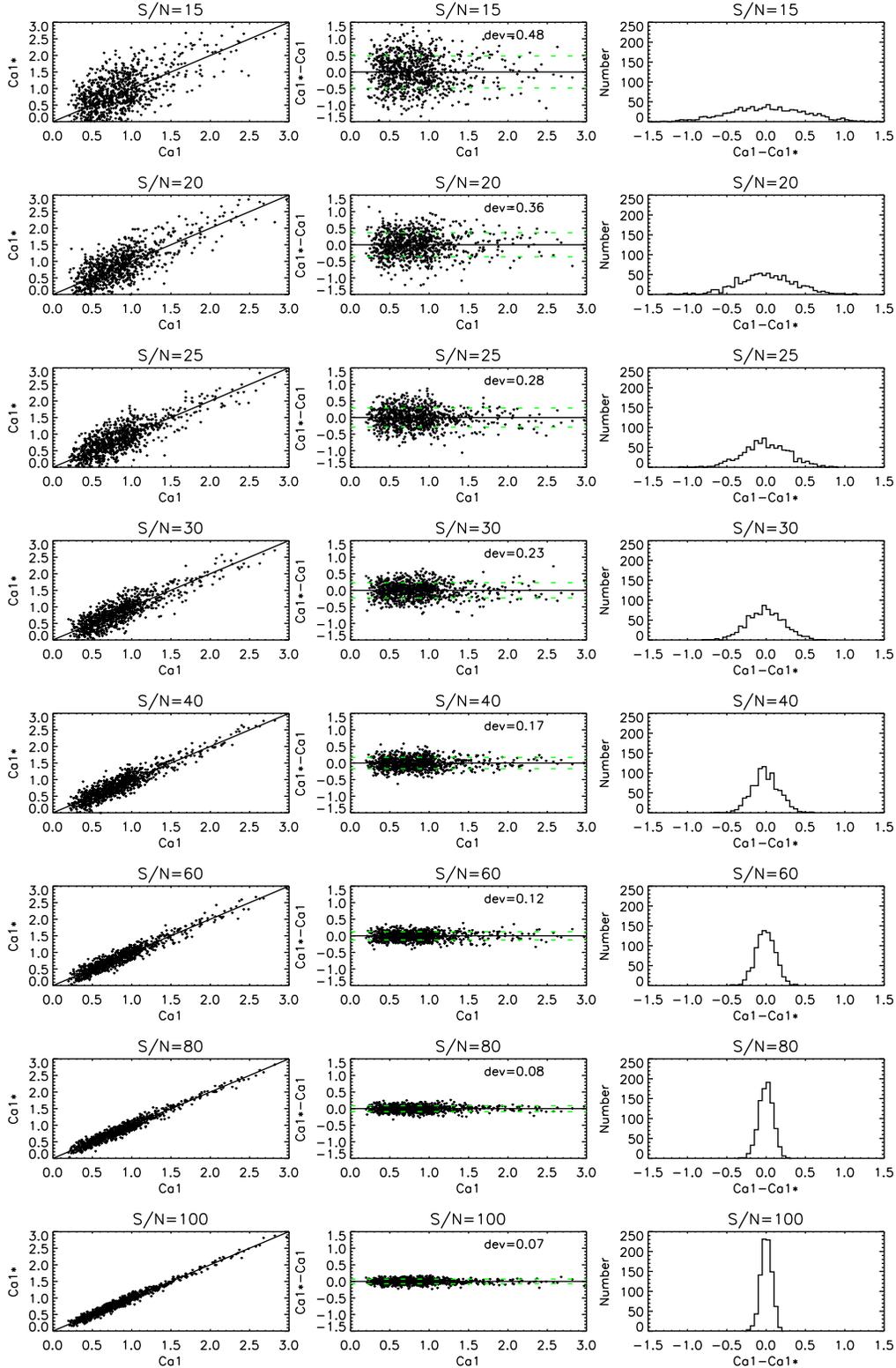}
 \caption[]{Sensitivity of Ca1 index to SNRs. Ca1* represents indices from synthetic spectra with no noise while Ca1 means indices from noise-added synthetic spectra. The left pannel shows Ca1* versus Ca1. The middle pannel directly shows the differences between Ca1* and Ca1, and 'dev' written down in this pannel is the standard deviation of the scatter. The black line is the zero-level line while the two green dashed lines decribe 1-$\sigma$ level. The right pannel shows the histogram distribution of differences between Ca1* and Ca1. }
\end{figure*}

\begin{figure*}
\centering
 \includegraphics[scale=1.8]{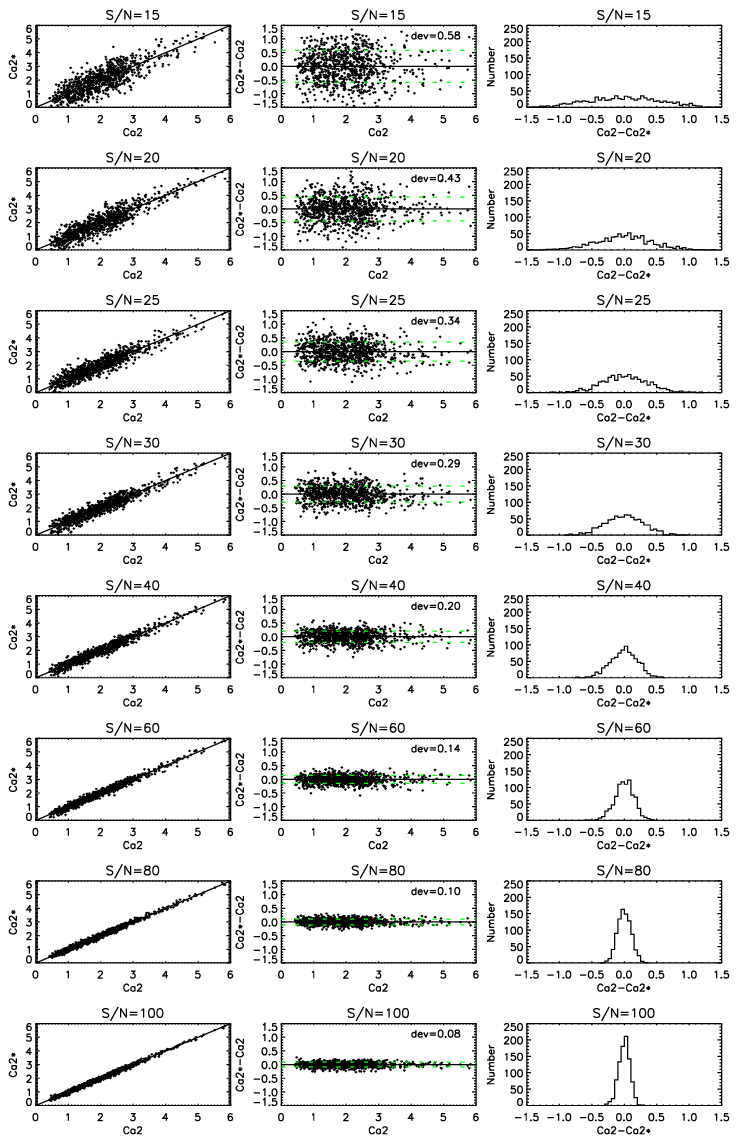}
 \caption[]{Sensitivity of Ca2 index to SNRs. Ca2* represents indices from synthetic spectra with no noise while Ca2 means indices from noise-added synthetic spectra. The left pannel shows Ca2* versus Ca2. The middle pannel directly shows the differences between Ca2* and Ca2, and 'dev' written down in this pannel is the standard deviation of the scatter. The black line is the zero-level line while the two green dashed lines decribe 1-$\sigma$ level. The right pannel shows the histogram distribution of differences between Ca2* and Ca2. }
\end{figure*}

\begin{figure*}
\centering
 \includegraphics[scale=1.8]{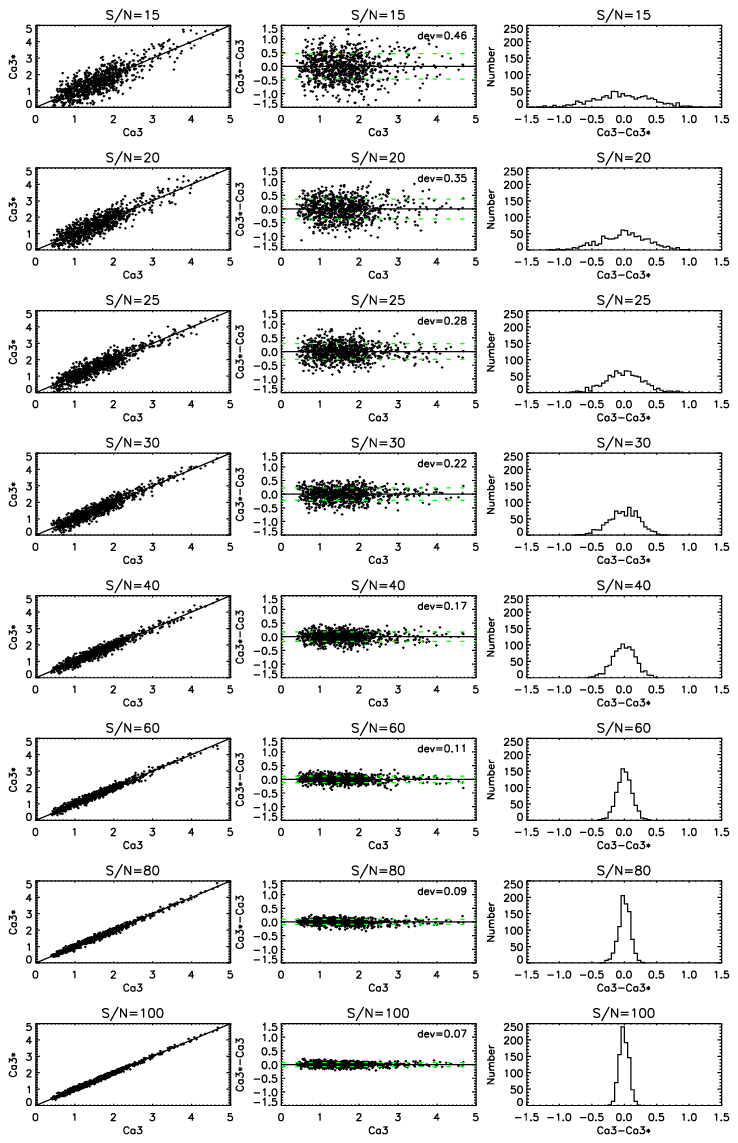}
 \caption[]{Sensitivity of Ca3 index to SNRs. Ca1* represents indices from synthetic spectra with no noise while Ca3 means indices from noise-added synthetic spectra. The left pannel shows Ca3* versus Ca3. The middle pannel directly shows the differences between Ca3* and Ca3, and 'dev' written down in this pannel is the standard deviation of the scatter. The black line is the zero-level line while the two green dashed lines decribe 1-$\sigma$ level. The right pannel shows the histogram distribution of differences between Ca3* and Ca3. }
\end{figure*}

\begin{figure*}
\centering
 \includegraphics[scale=1.8]{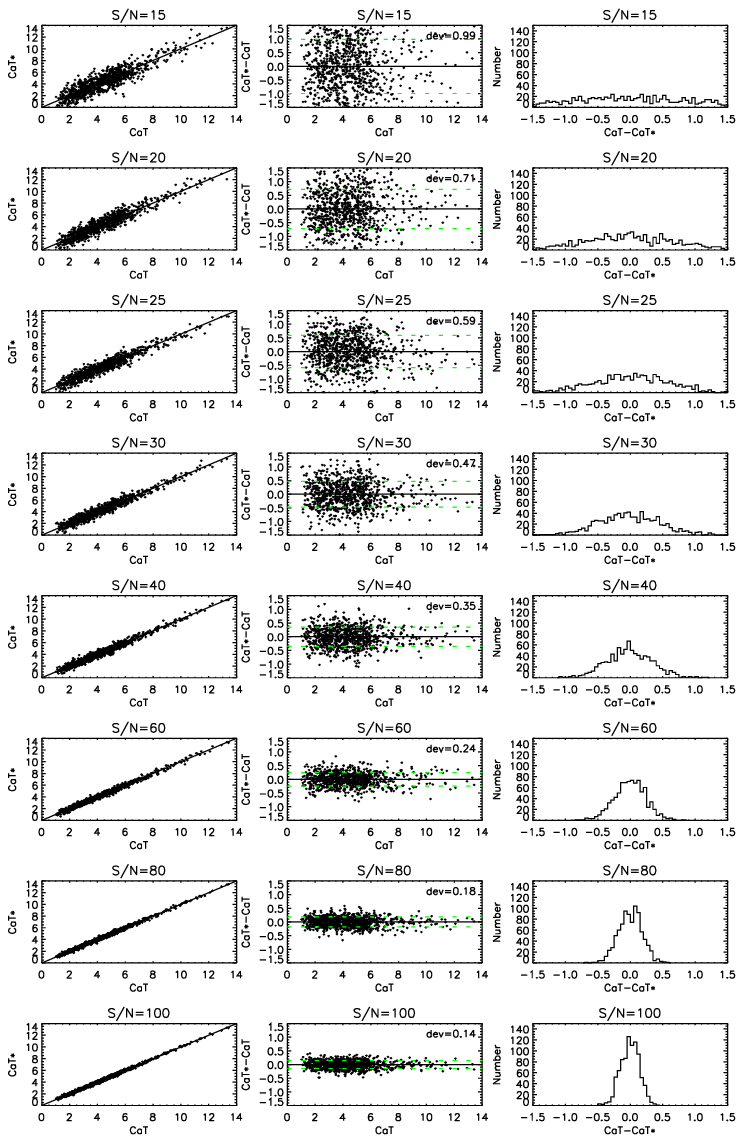}
 \caption[]{Sensitivity of CaT index to SNRs. CaT* represents indices from synthetic spectra with no noise while CaT means indices from noise-added synthetic spectra. The left pannel shows CaT* versus CaT. The middle pannel directly shows the differences between CaT* and CaT, and 'dev' written down in this pannel is the standard deviation of the scatter. The black line is the zero-level line while the two green dashed lines decribe 1-$\sigma$ level. The right pannel shows the histogram distribution of differences between CaT* and CaT. }
\end{figure*}




\begin{thebibliography}{99}
\bibitem[\protect\citeauthoryear{Abazajian et al.}{2009}]{b19}Abazajian K. N., et al., 2009, ApJS, 182, 543
\bibitem[\protect\citeauthoryear{Aihara et al.}{2011}]{b21}Aihara H., et al., 2011, ApJS, 193, 29
\bibitem[\protect\citeauthoryear{Burstein et al.}{1984}]{b10}Burstein D., Faber S. M., Gaskell C. M., Krumm N., 1984, ApJ, 287, 586
\bibitem[\protect\citeauthoryear{Burstein et al.}{1986}]{b12}Burstein D., Faber S. M., Gonzalez J. J., 1986, AJ, 91, 5
\bibitem[\protect\citeauthoryear{Carter et al.}{1986}]{b15}Carter D., Visvanathan N., Pickles A. J., 1986, ApJ, 311, 637
\bibitem[\protect\citeauthoryear{Castelli $\&$ Kurucz}{2003}]{b24}Castelli F., Kurucz R. L., 2003, in IAU Symp. 210, Modelling of Stellar Atmosphere, ed. N. E. Piskunov, W. W. Weiss, D. F. Gray(San Francisco, CA:ASP), A20
\bibitem[\protect\citeauthoryear{Cenarro et al.}{2001}]{b2}Cenarro A. J., Cardiel N., Gorgas J., Peletier R. F., Vazdekis A., Prada F., 2001, MNRAS, 326, 959
\bibitem[\protect\citeauthoryear{Diaz et al.}{1989}]{b1}Diaz A. I., Terlevich E., Terlevich R., 1989, MNRAS, 239, 325
\bibitem[\protect\citeauthoryear{Faber et al.}{1985}]{b11}Faber S. M., Friel E. D., Burstein D., Gaskell C. M., 1985, ApJS, 57, 711
\bibitem[\protect\citeauthoryear{Franchini et al.}{2010}]{b3}Franchini M., Morossi C., Marcantonio P. D., Malagnini M. L., Chavez M., 2010, ApJ, 719, 240
\bibitem[\protect\citeauthoryear{Gorgas et al.}{1993}]{b13}Gorgas J., Faber S. M., Burstein D., Gonzalez J. J., Courteau S., Prosser C., 1993, ApJS, 86, 153
\bibitem[\protect\citeauthoryear{Gray et al.}{1994}]{b22}Gray R. O., Corbally C. J., 1994, AJ, 107, 742
\bibitem[\protect\citeauthoryear{Jones et al.}{1984}]{b4}Jones J. E., Alloin D. M., Jones B. J. T., 1984, ApJ, 283, 457
\bibitem[\protect\citeauthoryear{Jorgensen et al.}{1992}]{b17}Jorgensen U. G., Carlsson M., Johnson H. R., 1992, A\&A, 254, 258
\bibitem[\protect\citeauthoryear{Luo et al.}{2008}]{b20}Luo A L., Wu Y., Zhao J., Zhao G., 2008, in Society of Photo-Optical Instrumentation Engineers (SPIE) Conference Series, vol.7019
\bibitem[\protect\citeauthoryear{Schwarzschild et al.}{1913}]{b26}Schwarzschild K., Eberhard G., 1913, ApJ, 38, 292
\bibitem[\protect\citeauthoryear{Serven et al.}{2005}]{b24}Serven J., Worthey G., Briley M. M., 2005, ApJ, 627, 754
\bibitem[\protect\citeauthoryear{Trager et al.}{1998}]{b14}Trager S. C., Worthey G., Faber S. M., Burstein D., Gonzalez J. J., 1998, ApJS, 116, 1
\bibitem[\protect\citeauthoryear{Valdes et al.}{2004}]{b6}Valdes F., Gupta R., Rose J. A., Singh H. P., Bell D. J., 2004, ApJS, 152, 251
\bibitem[\protect\citeauthoryear{Wilson $\&$ Vainu Bappu}{1957}]{b27}Wilson O. C., Vainu Bappu M. K., 1957, ApJ, 125, 661
\bibitem[\protect\citeauthoryear{Worthey et al.}{1994}]{b5}Worthey G., Faber S. M., Gonzalez J. J., Burstein D., 1994, ApJS, 94, 687
\bibitem[\protect\citeauthoryear{Worthey et al.}{2011}]{b25}Worthey G., Ingermann B. A., Serven J., 2011, ApJ, 729, 148
\bibitem[\protect\citeauthoryear{Wu et al.}{2011}]{b23}Wu Y., Singh H. P., Prugniel P., Gupta R., Koleva M., 2011, A\&A, 525, A71
\bibitem[\protect\citeauthoryear{York et al.}{2000}]{b18}York D. G., et al., 2000, AJ, 120, 1579
\bibitem[\protect\citeauthoryear{Zhou et al.}{1991}]{b16}Zhou X., 1991, A\&A, 248, 367





\end{thebibliography}
\end{document}